\documentclass{KapProc} 

\let\footnote\savefootnote
\let\footnotetext\savefootnotetext 
\let\footnoterule\savefootnoterule 

\kluwerbib

\startauthorindex

\usepackage{graphicx}

\begin{document}



\articletitle[Spatial Distributions of Multiple Dust Components 
in the PPN/PN Dust Shells]{%
Spatial Distributions of \\
Multiple Dust Components \\
in the PPN/PN Dust Shells}

\chaptitlerunninghead{Multiple Dust Components in the PPN/PN Dust Shells}

\author{Toshiya Ueta, Angela K. Speck, and Margaret Meixner}
\affil{University of Illinois at Urbana-Champaign}
\email{ueta@astro.uiuc.edu, akspeck@astro.uiuc.edu, meixner@astro.uiuc.edu}

\author{Aditya Dayal}
\affil{IPAC/JPL}
\email{adayal@ipac.caltech.edu}

\author{Joseph L. Hora and Giovanni Fazio}
\affil{Harvard-Smithsonian Center for Astrophysics}
\email{jhora@cfa.harvard.edu, gfazio@cfa.harvard.edu}

\author{Lynne K. Deutsch}
\affil{Boston University}
\email{deutschl@bebop.bu.edu}

\author{William F. Hoffmann}
\affil{Steward Observatory/University of Arizona}
\email{whoffmann@as.arizona.edu}


\begin{abstract}
We investigate spatial distributions of specific dust 
components in the circumstellar shells of a 
proto-planetary nebula candidate, \inx{HD 179821}, 
and a planetary nebula, \inx{BD$+$30$^{\circ}$3639},
by means of spectral imaging.
With high-resolution ground-based images and
{\sl ISO} spectra in the mid-infrared, we can derive 
``dust feature only'' maps
by subtracting synthesized continuum maps
from the observed images at the feature 
wavelength.
Such spatially detailed information will help
to develop models for
the evolution of dust grains around evolved stars.
\end{abstract}


\section{Introduction}

With {\sl ISO\/} discoveries of a wealth of infrared (IR) 
emission features from evolved stars (e.g., 
Waters et al.\ 1996; Waters et al.\ 1998) 
and laboratory measurements of IR spectra for candidate dust 
species (Speck 1998 for a compilation of references), 
we are now able to identify individual circumstellar 
dust components.

Using such detailed information 
to interpret high-resolution ground-based mid-IR images
we can spatially separate dust emission due to
different species by means of spectral imaging.
This can be done by
(1) identify an emission feature and nearby continuum bands,
(2) construct a color temperature map from the continuum images,
(3) synthesize a continuum image at the feature wavelength from
the temperature map taking into account diffraction effects,
and
(4) subtract the synthesized continuum-only map from the 
observed feature band image.

\section{Dust Feature Map Construction}

Fig. 1 shows mid-IR images and spectra of
an oxygen-rich proto-planetary nebula candidate,
\inx{HD 179821}~(\inx{IRAS 19114+0002}, \inx{AFGL 2343}).
The images indicate the presence of a dust torus 
surrounding the central star, and 
the spectra reveal both amorphous 
and crystalline
silicate features 
(Waters et al.\ 1996).
We can isolate dust emission at 10.3 and 
$18 \mu$m by subtracting continuum 
emission maps synthesized
using 8.8, 12.5, and $20.6 \mu$m 
band images as reference continuum.

\begin{figure}[ht]
\includegraphics[width=17.75pc]
{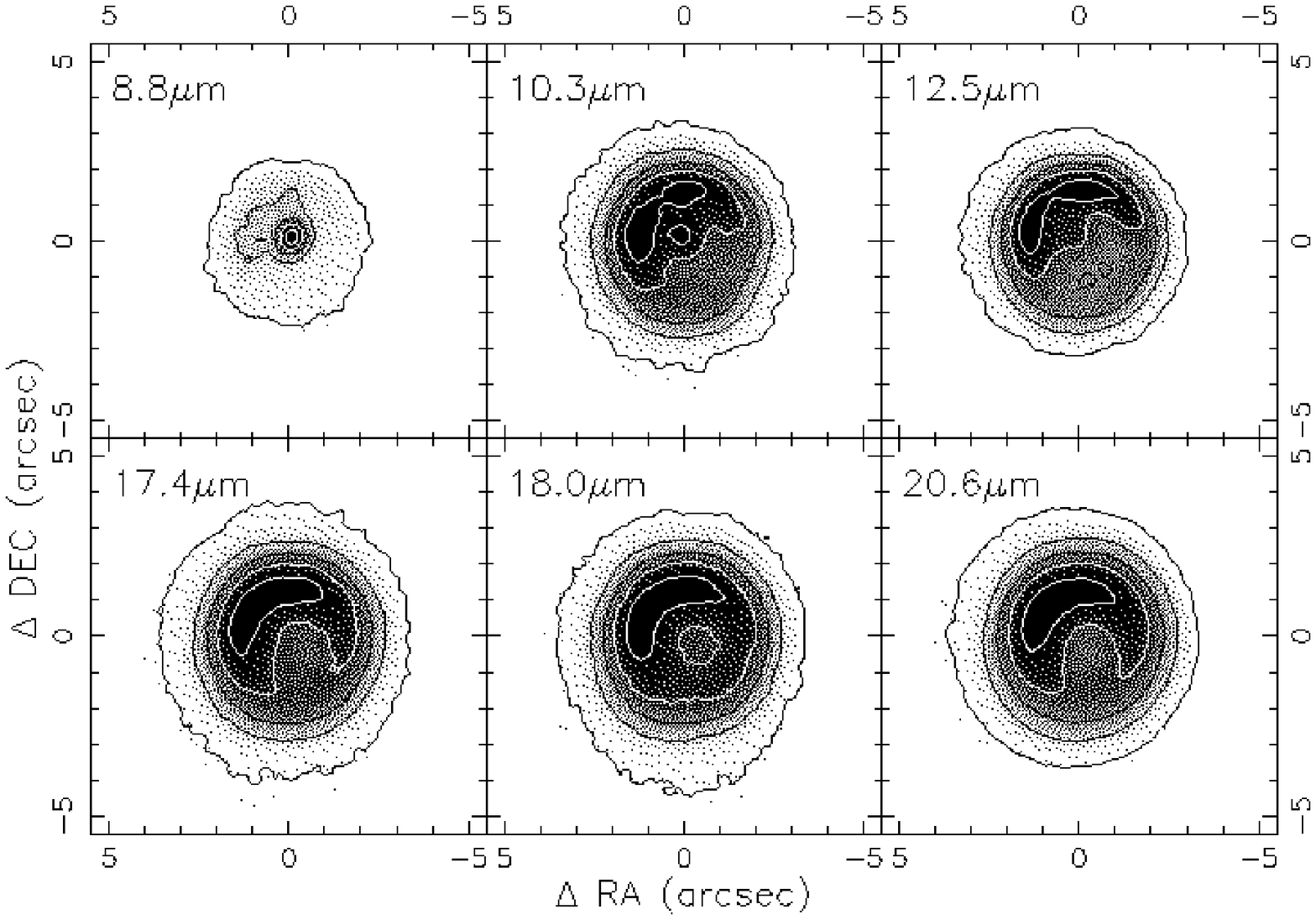}
\includegraphics[width=9.25pc]
{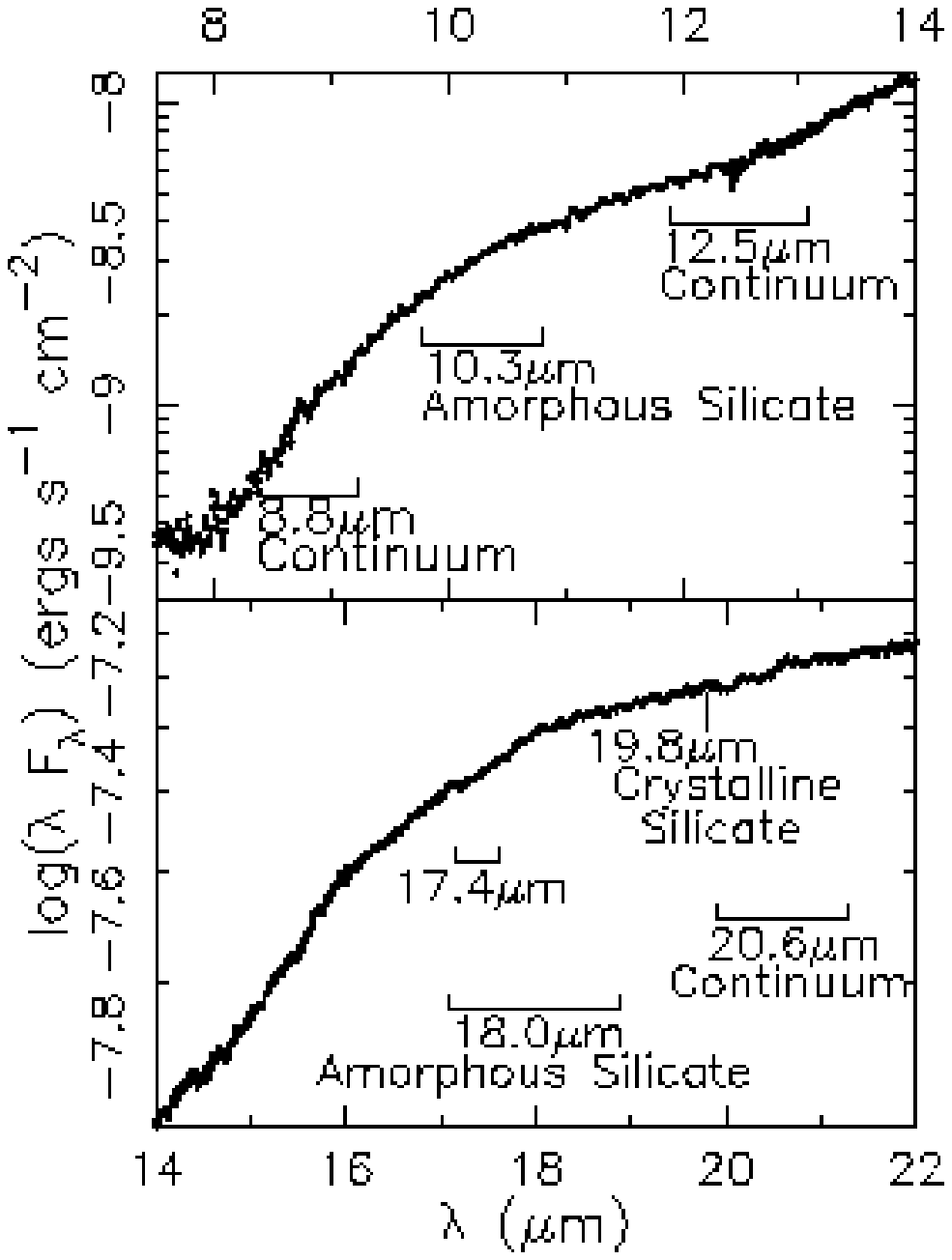}
\caption{MIRAC3/IRTF images (left: contours are
90, 70, 50, 30, and 10\% of the peak) and 
{\sl ISO\/} spectra (right; Waters et al.\/ 1996) 
of \protect\inx{HD 179821}
Band/filter locations are indicated in the spectra.}
\end{figure}

\inx{BD$+$30$^{\circ}$3639} (\inx{IRAS 19327$+$3024}) 
is a carbon-rich planetary nebula, in which 
crystalline silicates have been detected
(Waters et al.\ 1998).
Mid-IR images (Fig. 2) show a box-like elongation which 
is also seen in optical emission line images
(e.g., Harrington et al.\ 1997).

\begin{figure}[h]
\includegraphics[width=17.75pc,height=1.8in]
{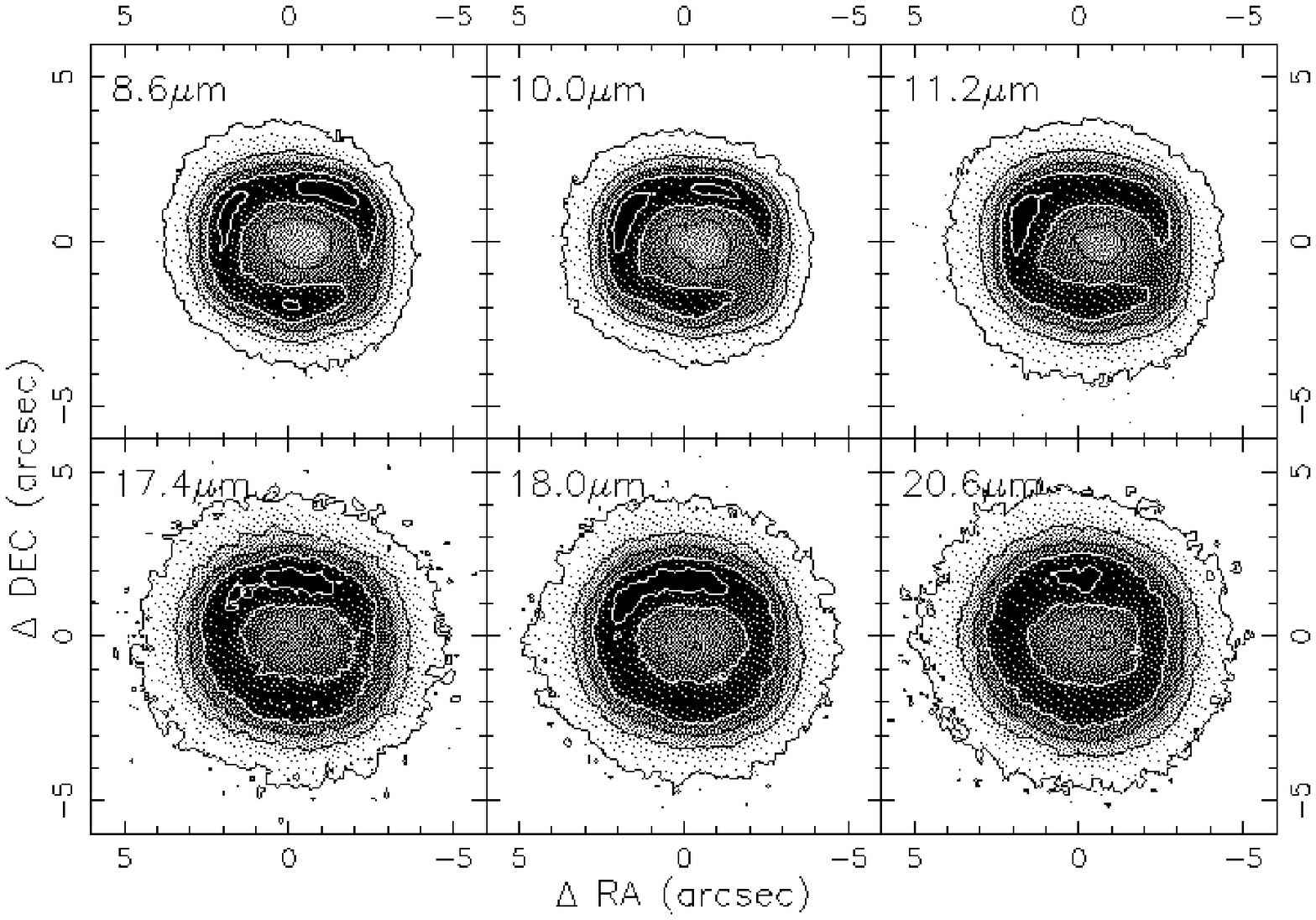}
\includegraphics[width=9.25pc,height=1.8in]
{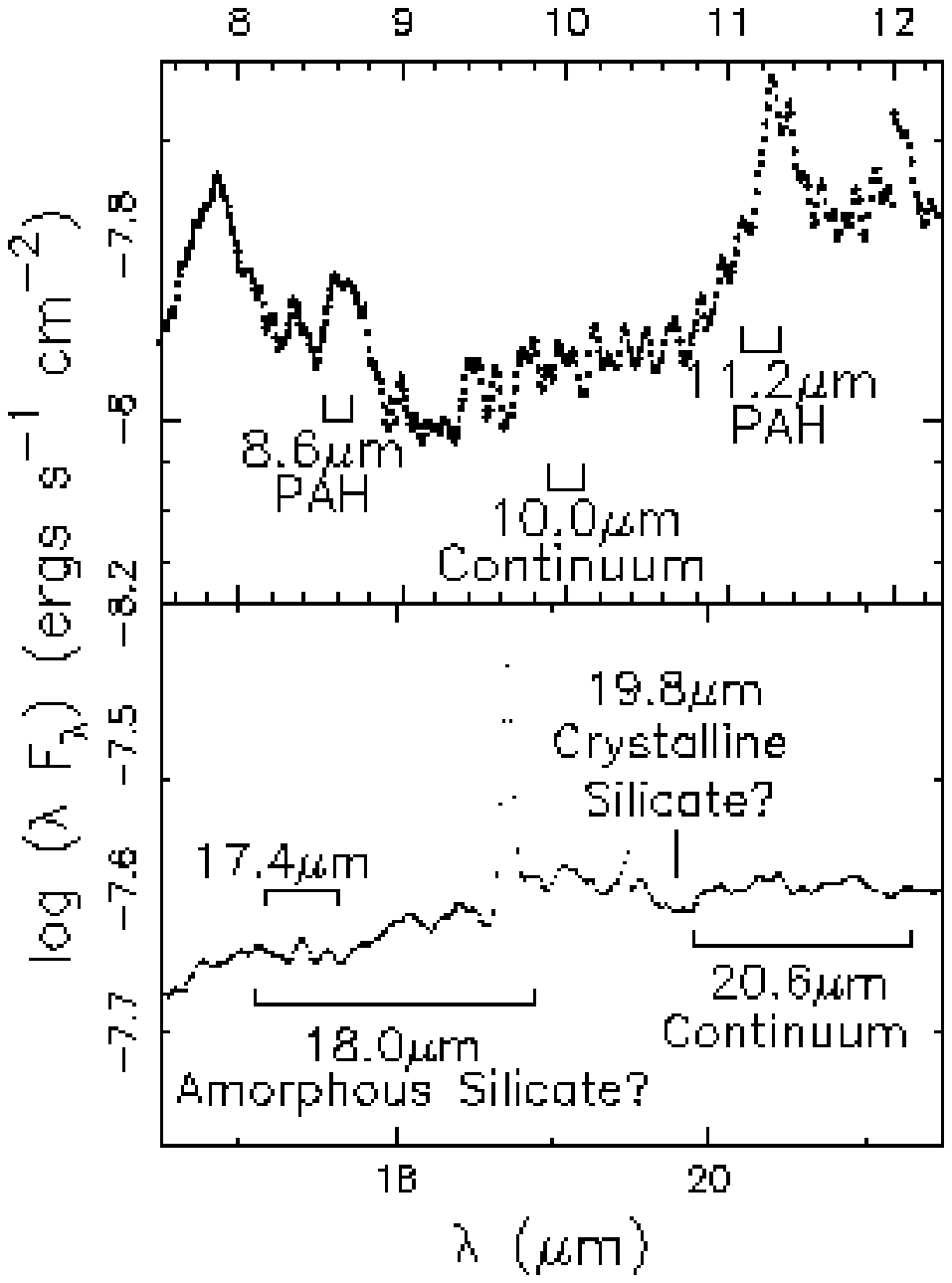}
\caption{MIRAC3/IRTF images (left) and 
{\sl ISO\/} spectra (right; Waters et al.\/ 1996) 
of \protect\inx{BD$+$30$^{\circ}$3639}~in the mid-IR.
Band/filter locations are indicated in the spectra.}
\end{figure}

While
polyaromatic hydrocarbon (PAH; e.g., Allamandola et al.\ 1999)
emission at 8.6 and $11.2 \mu$m can be isolated
by using 10 and $20.6 \mu$m continuum band images (Fig. 4),
we were unable to isolate crystalline silicate at $19.8 \mu$m
because of the weak feature and apparent lack
of clean continuum around $18 \mu$m. 

\section{Results and Implications}

\begin{figure}[b]
\includegraphics[width=13.5pc,height=2.in]
{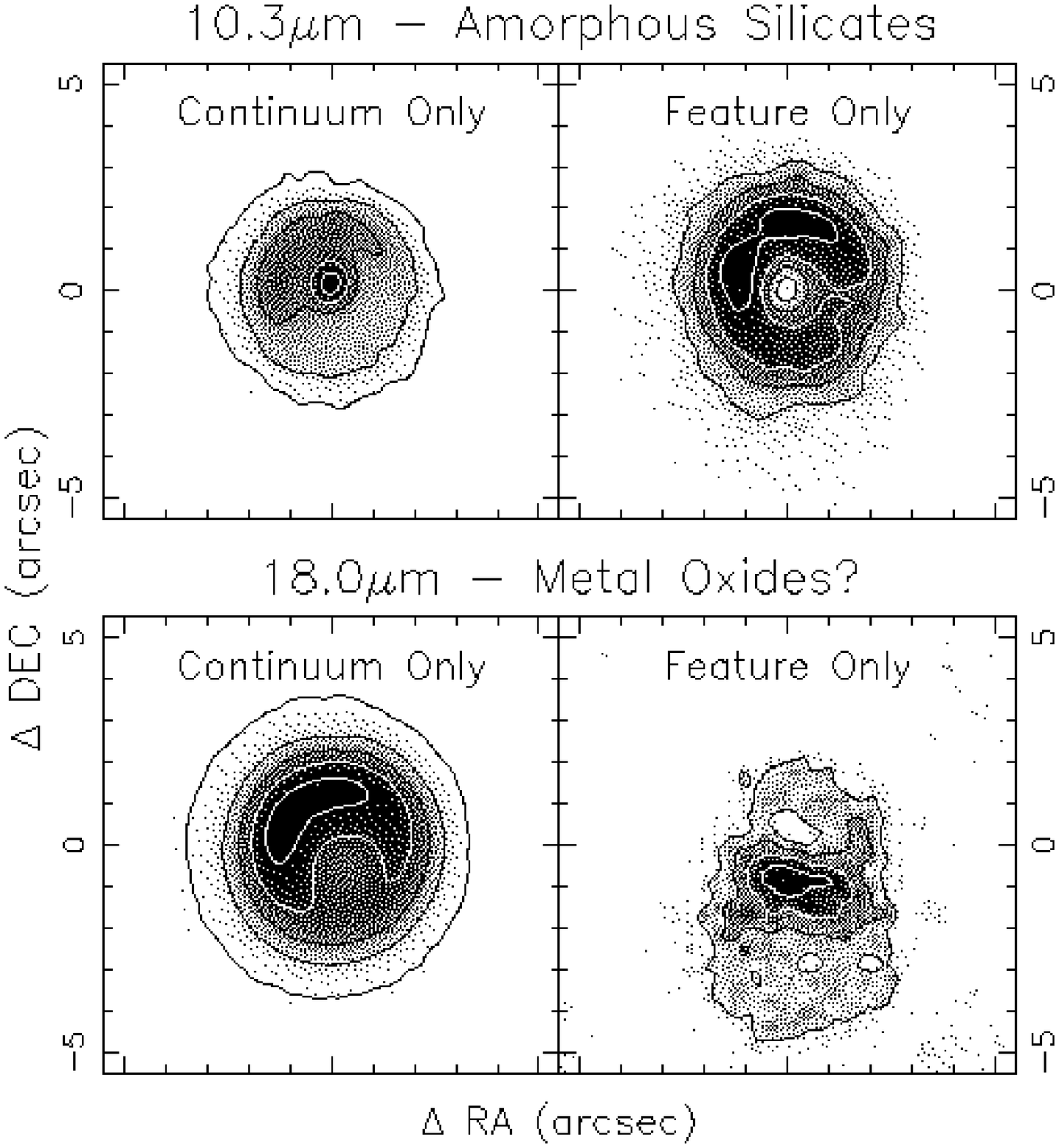} 
\includegraphics[width=13.5pc,height=2.in]
{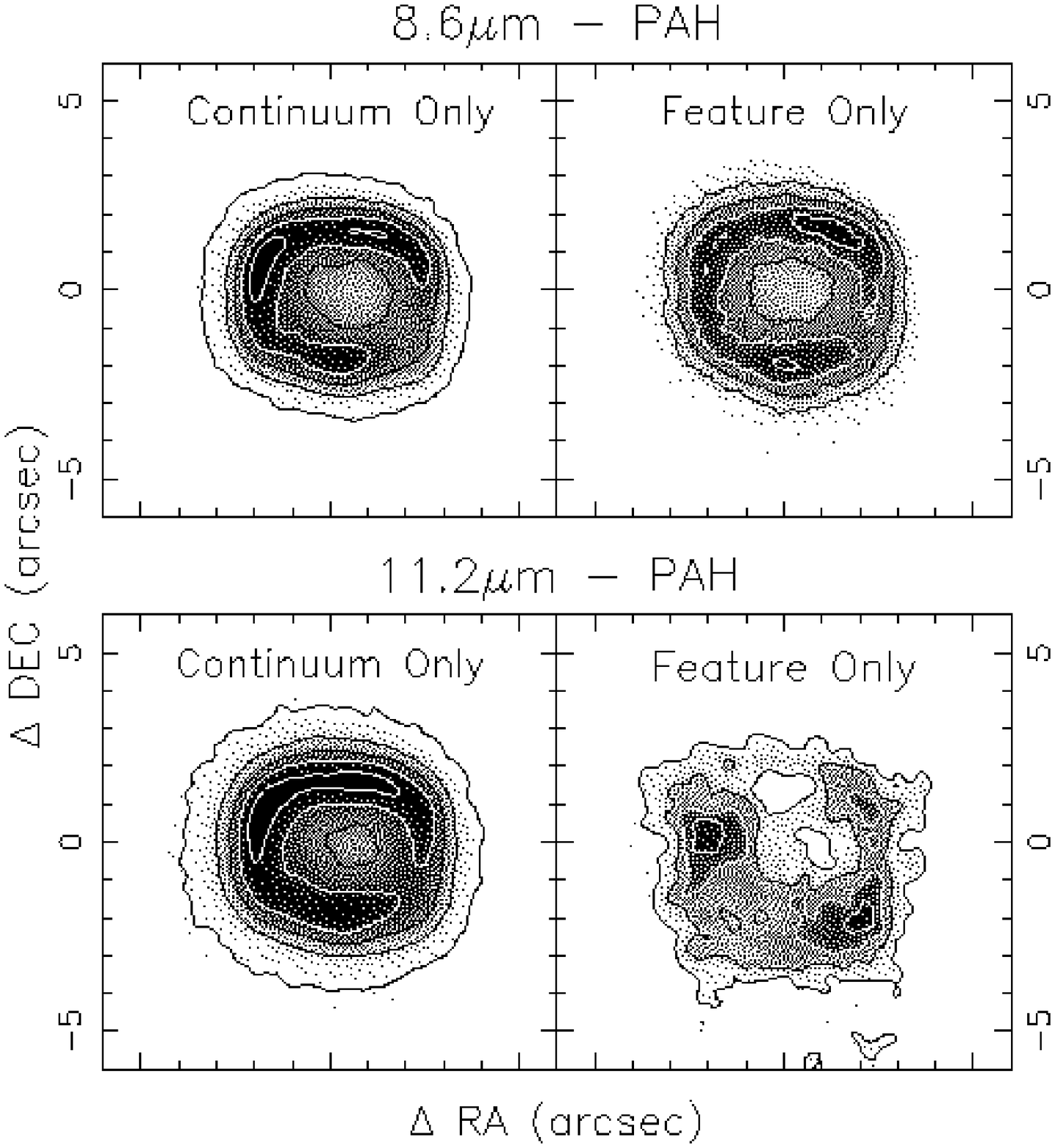} 
\dblcaption{Continuum/feature only images at 10.3 and $18.0 \mu$m
for \protect\inx{HD 179821}.}
{Continuum/feature only images at 8.6 and $11.2 \mu$m
for \protect\inx{BD$+$30$^{\circ}$3639}.}
\end{figure}

\newpage

\noindent
\begin{minipage}{18pc}
\hspace{1pc}Fig. 3 shows that the 10.3 and 18.0$\mu$m
bands have different spatial distributions, indicating that
these emissions are due to different carrier species.
We thus need to find a dust species with an
emission feature at 18$\mu$m but not at 10.3$\mu$m
(i.e., not silicates).
Magnesium iron oxides are such species as shown in the 
spectrum
(Henning et al.\ 1995).\\
\end{minipage}
\hfill
\begin{minipage}{10pc}
\includegraphics[angle=-90,width=10pc]
{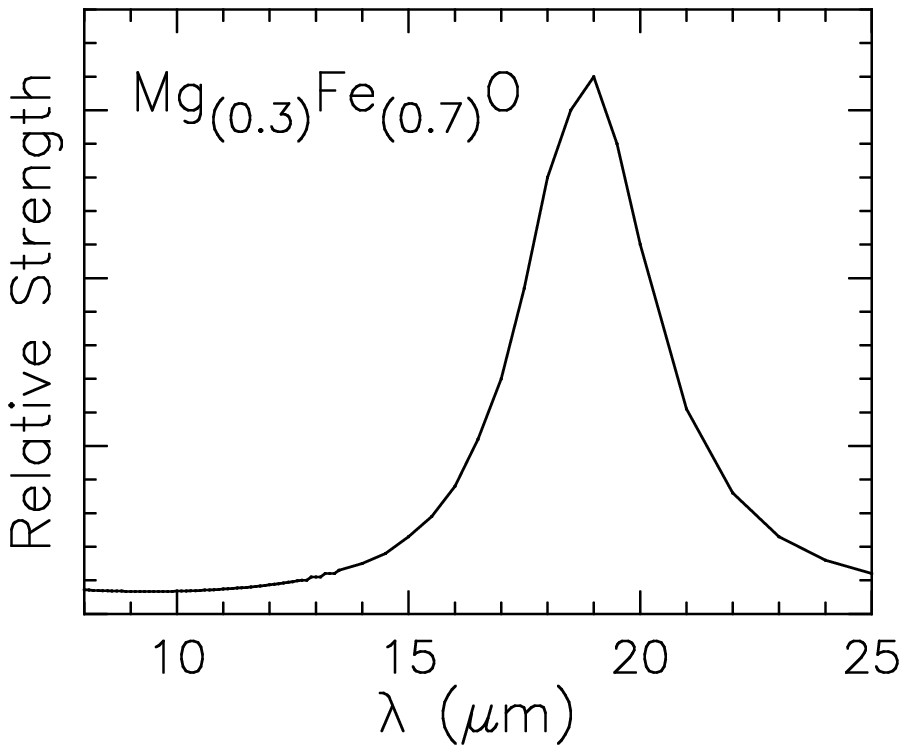} 
\end{minipage}

PAH emissions at 8.6 and 11.2$\mu$m show
drastically different spatial distributions
in Fig. 4, implying
that these PAH bands may have been
originated differently
(e.g., Beintema et al.\ 1996, Molster et al.\ 1996).\\

\hfill
\begin{minipage}{9pc}
 {\bf The 8.6$\mu$m feature}\\
$\bullet$ slightly compact\\
$\bullet$ ionized PAHs\\
$\bullet$ in-plane C-H mode\\
\end{minipage}
\hfill
\begin{minipage}{2pc}
 $\Longleftrightarrow$
\end{minipage}
\hfill
\begin{minipage}{12pc}
{\bf The 11.2$\mu$m feature}\\
$\bullet$ slightly extended\\
$\bullet$ neutral PAHs\\
$\bullet$ out-of-plane C-H mode\\
\end{minipage}

\noindent
Observation of the differences listed above
suggests that the difference between the two
bands may arise from the availability of hydrogen, either
in the form of an atom or an out-of-plane C-H stretch.
The former implies the difference in the ionization 
state of PAHs
(i.e., the 8.6$\mu$m emission is from the ionized 
shell and the 11.2$\mu$m emission is from the 
photo dissociation region),
while the latter suggests different structure of PAHs
(i.e., the 8.6$\mu$m emission is from PAH clusters
and the $11.2\mu$m emission is from chain- or
sheet-like PAHs).

Careful selection/tailoring of filters will make
spectral imaging more effective.
Determining the spatial distribution of specific dust 
species
will allow the development of models for the evolution of dust 
around evolved stars. 
Understanding dust evolution benefits other aspects of the
physics involved such as grain nucleation and growth,
the chemical evolution of the stellar atmospheres,
and the nucleosynthesis of stars.




\end{document}